\newcommand{\be}{\begin{equation}}\newcommand{\ee}{\end{equation}}
\newcommand{\bea}{\begin{eqnarray}}\newcommand{\eea}{\end{eqnarray}}
\newcommand{\nn}{\nonumber \\}\newcommand{\p}[1]{(\ref{#1})}

\newcommand\Hb{\overline H}
\newcommand\Phib{\overline \Phi}

\newcommand\D{{\cal D}}
\newcommand\Psib{\overline{\Psi}}
\newcommand\Db{\overline{\cal D}}
\newcommand\DDb{\left[ {\cal D},{\overline{\cal D}}\right]}
\documentstyle[12pt]{article}
\def\theequation{\arabic{section}.\arabic{equation}}
\begin{document}
\setcounter{page}0
\renewcommand{\thefootnote}{\fnsymbol{footnote}}
\thispagestyle{empty}
\begin{flushright}
April 1997 \\
SISSA 59/97/EP\\
solv-int/9705001
\vspace{2cm}\\ \end{flushright} 
\begin{center}
{\large\bf Hamiltonian structure and coset construction of the  
supersymmetric extensions of $N=2$ KdV hierarchy}
\vspace{0.5cm} \\
L. Bonora\footnote{E-mail: bonora@frodo.he.sissa.it}${}^a$
and S. Krivonos\footnote{E-mail: krivonos@thsun1.jinr.dubna.su}${}^b$ 
  \vspace{0.5cm} \\
{\it ${}^{(a)}$ International School for Advanced Studies (SISSA/ISAS)} \\
{\it Via Beirut 2, 34014 Trieste, Italy} \\
{\it INFN, Sezione di Trieste}\\
{\it ${}^{(b)}$Bogoliubov  Laboratory of Theoretical Physics, JINR,\\
141980 Dubna, Moscow Region, Russia} \vspace{1.5cm} \\
{\bf Abstract}
\end{center}
A manifestly $N=2$ supersymmetric coset formalism is applied to analyse
the "fermionic" extensions of $N=2$ $a=4$ and $a=-2$ KdV hierarchies. 
Both these hierarchies can be obtained from a manifest $N=2$ coset 
construction. This coset is defined as the quotient of some local
but non-linear superalgebra by a $\widehat{U(1)}$ subalgebra.
Three superextensions of $N=2$ KdV hierarchy are proposed, among which one
seems to be entirely new.
\newpage
\renewcommand{\thefootnote}{\arabic{footnote}}
\setcounter{footnote}0
\setcounter{equation}0\section{Introduction}
The search for $N=2$ and $N=4$ extensions of known hierarchies of integrable 
equations, has brought new impulse to the research in the field of integrable
systems. Not only are these extensions characterized by beautiful 
supersymmetric second hamiltonian structures, but, what is more important, 
they seem to be endowed with an organizing capacity, that allows them to 
encompass different cases under a unique structure. Newertherless, in many
cases adding some additional fields or even changing the statistics of
the fields into the game -- which are typical ways of getting new hierarchies 
from old ones -- may preserve integrability but make the second
Hamiltonian structure non-local \cite{{DG},{BS}}.
The purpose of this letter is to demonstrate on the simplest possible examples
that the systems with non-local second Hamiltonian structure are
subsystems of extended integrable systems with a {\it local} second
Hamiltonian structure and can be obtained from the latter via the 
coset approach. One can hope that this represents the general situation and 
that the fundamental integrable hierarchies possess local second
Hamiltonian structures.

We start from ``fermionic'' extensions of $N=2,a=4$ \cite{{IK},{DG}} and 
$N=2,a=-2$ \cite{DGI} KdV hierarchies which possess the same non--local
second hamiltonian structure. It turns out that the best way to unify these 
hierarchies is via a coset construction which `localizes' the second 
hamiltonian structure. We explicitly construct the corresponding
coset in section 2. In section 3 we show that the Dirac reduction
applied to the corresponding local but non-linear algebra reproduces the
non-local algebra that constitutes the original second Hamiltonian structure.
In section 4, as the first step to analyse these extended systems,
we search for integrable systems that have 
a subalgebra of our non-linear algebra as their second Hamiltonian
structure. Two solutions correspond to supersymmetric extensions of
already known integrable hierarchies,
but we also introduce the first flows and hamiltonians of an
$N=2$ hierarchy, which does not seem to have any correspondence among the 
cases studied so far.

\setcounter{equation}0
\section{"Fermionic" extensions of the N=2 KdV in the coset approach}

\subsection {\bf Preliminaries: N=4 KdV hierarchy}

The $N=4$ supersymmetric KdV equation has been constructed in \cite{DI}.
In terms of $N=2$ superfields it can be written as the coupled
system of equations for the general bosonic superfield $J$ and for a
pair of chiral/anti-chiral bosonic superfields $\Phi$/$\Phib$.
Explicitly, the first two non-trivial flows (second and third) have
the following form
\bea 
\frac{\partial J}{\partial t_2} & = & -\DDb J'-4JJ'-
     \left( \Phi \Phib \right){}' \; , \nn
\frac{\partial \Phi}{\partial t_2} & = & -\Phi{}''+ 
   4\D\Db \left( J \Phi \right) \; , \;
\frac{\partial \Phib}{\partial t_2}
  =  \Phib{}''+ 
   4\Db\D \left( J \Phib \right)  \label{n4kdv2}
\eea
and 
\bea 
\frac{\partial J}{\partial t_3} & = & J'''+3\left(\DDb J\; J\right)'+
  \frac{3}{2}\left(\DDb J^2\right)'+4\left( J^3\right)'+
   \frac{3}{2}\left(\D\Phib\Db\Phi\right)'+6\left( J\Phi\Phib\right)' \;,\nn
\frac{\partial \Phi}{\partial t_3} & = & \Phi{}'''+ 
   6\D\Db \left( \D (J\Db\Phi )-2 J^2 \Phi \right) ,\; 
\frac{\partial \Phib}{\partial t_3} =  \Phib{}'''-
   6\Db\D \left( \Db (J\D\Phib ) -2J^2 \Phib \right) \; , \label{n4kdv3}
\eea
where prime means differentiation over space coordinate $\left\{z\right\}$, 
and $\D$ and $\Db$ are the $N=2$ supersymmetric Grassmann covariant 
derivatives
\be \label{defDDB}
\D=\frac{\partial}{\partial\theta}-
  \frac{\bar\theta}{2}\frac{\partial}{\partial z}\; , \;
\Db=\frac{\partial}{\partial{\bar\theta}}-
  \frac{\theta}{2}\frac{\partial}{\partial z}\; , \;
\D^2=\Db{}^2=0 \; ,\; \left\{ \D ,\Db \right\}=-\partial_z\equiv -\partial \;.
\ee
The superfields $\Phi$ and $\Phib$ are constrained to be chiral-anti-chiral
correspondingly:
\be\label{chir}
\D \Phi = \Db \Phib =0 \; .
\ee
The integrability of the $N=4$ KdV hierarchy \p{n4kdv2},\p{n4kdv3} has been
proved in \cite{{IK},{DG}} by explicit construction of the Lax operator 
and the Lax equation
\be\label{laxDG}
L=\D\Db +\D\Db\partial^{-1}\left( J+ \Phib\partial^{-1}\Phi\right)
 \partial^{-1} \D\Db \; , \quad
\frac{\partial L}{\partial t_k}=-\left[ L^k_{\geq 1},L\right] .
\ee
where the subscript $\left\{ \geq 1 \right\}$ indicates the strictly 
differential part of the pseudodifferential operator. 
One can easily check that the second and third flow equations for the Lax 
operator \p{laxDG} coincide with \p{n4kdv2}-\p{n4kdv3}. 
The infinite set of conserved currents ${\cal H}_k$ for the $N=4$ KdV 
hierarchy is given by the following expression 
\be
H_k \equiv \int dzd\theta d{\bar\theta} {\cal H}_k=
        \int dzd\theta d{\bar\theta}  \;\mbox{Res } \left( L^k\right)
\ee
where the residue is defined as the coefficient before $\DDb \partial^{-1}$.

The main attractive feature of the $N=4$ KdV hierarchy is that it
possesses the classical "small" $N=4$ SCA (with $su(2)$ affine
subalgebra) as the second Hamiltonian
structure \cite{DI} (actually this was the reason why it was called
$N=4$ KdV hierarchy). Indeed, if we define the PB's for
the "small" $N=4$ SCA as
\bea
\left\{ J(1),J(2) \right\}& = & -\left( J\partial +\partial J+
 \D J\Db+\Db J\D+ \frac{1}{2}\partial \DDb \right) \delta (1,2) \; ,\nn
\left\{ J(1),\Phi (2) \right\} & = & \left( \Phi\Db\D +\Db\Phi\D\right)
            \delta (1,2) \; , \nn
\left\{ J(1),\Phib (2) \right\} & = & \left( \Phib\D\Db +\D\Phib\Db\right)
            \delta (1,2) \; , \nn
\left\{ \Phi(1),\Phib (2) \right\}& = & -2\left( \partial\D\Db +
                2\D J\Db\right) \delta (1,2) \; ,\label{n4sca}
\end{eqnarray}
where $\delta (1,2)$ is the $N=2$ superspace delta-function
\begin{equation}
\delta (1,2) = (\theta_1-\theta_2)({\bar\theta}_1-{\bar\theta}_2)
    \delta (z_1-z_2) \; ,
\end{equation}
and the differential operators in the r.h.s are evaluated  at the point $Z_1$
and the derivatives in the r.h.s are assumed to act freely to the right,
then the flow equations for the $N=4$ KdV hierarchy can be written
in hamiltonian form as follows
\be
\frac{\partial }{\partial t_k}\left(
   \begin{array}{c} J \\ \Phi \\ \Phib \end{array}\right)=
 \left\{ \left(\begin{array}{c} J \\ \Phi \\ \Phib \end{array}\right),
  \right. \left. H_k \right\} \; .
\ee
The hamiltonians $H_2,H_3$  which give rise to the second and third flows of 
$N=4$ KdV hierarchy \p{n4kdv2}-\p{n4kdv3} have the following explicit form
\bea 
H_2 & = & - \int dzd\theta d{\bar \theta} \left( J^2-
   \frac{1}{2}\Phib\Phi \right) \; , \nn
H_3 & = &  \int dzd\theta d{\bar \theta} \left(
 \DDb J\; J +\frac{4}{3}J^3+2J\Phi\Phib +\frac{1}{2}\Phi'\Phib \right).
\eea

\subsection {\bf "Fermionic" extensions of N=2, a=4,-2 KdV hierarchies}

A close inspection of the Lax operator for the $N=4$ KdV hierarchy \p{laxDG}
gives a hint as to how to construct integrable hierarchies which
have $N=2$ $a=4$ KdV in the reduction limit $\Phi=\Phib=0$.
Indeed, the superfields $\Phi$ and $\Phib$
appear in the Lax operator only in pairs. So one may wonder whether
the Lax equation  \p{laxDG} will be consistent if we
change the statistics of the superfields $\Phi,\Phib$ so that they become
fermionic superfields or if we even add an arbitrary number of additional
superfields with arbitrary statistics, while keeping the same form of the
Lax operator
\be\label{laxDGe}
L=\D\Db +\D\Db\partial^{-1}\left( J+ 
  \sum_i \Phib^{i}\partial^{-1}\Phi^{i} +
  \sum_j \Psib^{j}\partial^{-1}\Psi^{j}\right)
 \partial^{-1} \D\Db \; , 
\ee
where $\Psib^{j},\Psi^{j}$ are fermionic chiral/anti-chiral superfields. 
The answer is positive, but the price we have to pay is non-locality
of the second Hamiltonian structure \cite{{DG},{BS}}, even in the case
of only one pair of fermionic superfields $(i=0,j=1)$. Let us note that 
for this special case the equations of the second and third flows have the same
form as their bosonic conterparts \p{n4kdv2},\p{n4kdv3}.

From the integrability point of view the Lax operator and Lax equation
give us the full information about system. However non-locality of the
second Hamiltonian structure is a somewhat unappealing property and we 
would rather prefer tp understand the origin of such non--locality. 
Therefore we rise the question: is it  
possible to construct some extended system which reproduce these ones
as particular cases when some additional fields are set to zero, and
are chararcterized by a {\it local} second Hamiltonian structure? 
If such extended systems exist they are certainly more general and 
more fundamental. In what follows we 
consider the simplest case with one pair of fermionic chiral-anti-chiral
superfields and we will show that it is indeed possible to construct
the extended system of the type just described.

The simplest possibility to construct a {\it local} algebra which contains
at least all superfields that feature in the fermionic version of the
equations \p{n4kdv2},\p{n4kdv3}, i.e. the bosonic $J$ together with the 
fermionic 
superfields $\Psi,\Psib$, is to add one additional pair of chiral-anti-chiral
fermionic superfields $H,\Hb$. This extended set of superfields forms 
a closed {\it local} although {\it non-linear} algebra  
\bea
\left\{ J(1),J(2) \right\}& = & -\left( J\partial +\partial J+
 \D J\Db+\Db J\D+ \frac{1}{2}\partial \DDb \right) \delta (1,2)  ,\nn
\left\{ J(1),\Psi (2) \right\}& = & \left( 2\Db\D\Psi+\D\Psi\Db+
 \Db\Psi\D-\Psi\partial \right) \delta (1,2)  ,\nn
\left\{ J(1),\Psib (2) \right\}& = & \left( 2\D\Db\Psib+\D\Psib\Db+
 \Db\Psib\D-\Psib\partial \right) \delta (1,2)  ,\nn
\left\{ J(1),H (2) \right\} & = &  H\Db\D \delta (1,2)  , \;
\left\{ J(1),\Hb (2) \right\}  =  \Hb\D\Db \delta (1,2)  , \nn
\left\{ H(1),\Hb (2) \right\}& = & \frac{1}{2}\D\Db \delta (1,2)  , \;
\left\{ H(1),\Psi (2) \right\} = \D\Psi \delta (1,2)  , \;
\left\{ H(1),\Psib (2) \right\} = -\D\Psib \delta (1,2)  , \nn
\left\{ \Hb(1),\Psi (2) \right\}& = & -\Db\Psi  \delta (1,2)   , \;
\left\{ \Hb(1),\Psib (2) \right\} = \Db\Psib\; \delta (1,2)  ,   \nn
\left\{ \Psi (1),\Psib (2) \right\}& = & 2\left( -\partial\D\Db-
   2\D J\Db+\Psi\Psib-4\D J\Hb -4JH\Db-8JH\Hb +2\D\Hb\D\Db \right. \nn
  & & -2\partial\D\Hb-2\partial H\Db-2H\partial\Db+2\Db H\D\Db+
      4\D\Hb\D\Hb+4H\Db H\Db \nn
  & & +8H\Hb\D\Hb+8H\Hb\Db H+4\D\Db H\Hb+4H\Db\D\Hb \nn
  & & \left. -8H\D\Db\Hb+4H\Hb\D\Db\right)\delta (1,2) \; , \label{newsca}
\end{eqnarray}
which we claim is the right second Hamiltonian structure for the fermionic
version of the system \p{n4kdv2},\p{n4kdv3} in the coset approach.

One can check that the Jacobi identities put additional constraints on
the superfields $\Psi,\Psib$ :
\be\label{constr}
\D\Psi = -2\; H\Psi \; , \; \Db\Psib=2\;\Hb\Psib \; .
\ee
The set of Poisson brackets \p{newsca} together with the covariant chirality
constraints \p{constr} form the closed algebra we are looking 
  for \footnote{To our best knowledge, this algebra first appeared in hidden 
form (as subalgebra of
some extended algebra in a special basis) in the Ref.\cite{AIS}.}.
This algebra has the structure of non-linear $W$ algebra due to the presence
of non-linear terms in the right hand side.

The algebra \p{newsca} possesses manifest $N=2$ supersymmetry. The same
is true also for the coset which is defined by quotienting it by the 
$\widehat{U(1)}$ subalgebra generated by $H,\Hb$. It is this coset that forms 
the
second Hamiltonian structure for the "fermionic" extension of $N=2, a=4$ KdV
hierarchy. 
Stated another way, it is possible to construct an infinite
number of Hamiltonians which have vanishing Poisson brackets
with respect to $H,\Hb$. Thus, at the level of equations of motion
one can consistently put $H=\Hb=0$ and remain with the equations for
the fields $J,\Psi,\Psib$. Moreover, after putting $H$ and $\Hb$ equal to
zero, the covariant chirality conditions \p{constr} will become the
ordinary chiral/anti-chiral ones
\be\label{constr1}
\D\Psi = \Db\Psib= 0 \; .
\ee
The Hamiltonian densities ${\cal H}_k$ in the $N=2$ superspace for the integer 
$k=1,2,\ldots $ are bosonic and have conformal dimension $k$. 
The explicit form of the first three Hamiltonians  is
\bea
H_1 & = & \int dzd\theta d{\bar \theta} \; J \; , \;\qquad
H_2  = \int dzd\theta d{\bar \theta} \left(  
      -J^2+\frac{1}{2}\Psib\Psi\right) \; \nn
H_3 & = & \int dzd\theta d{\bar \theta} \left( 
   J\DDb J +2J\Psi\Psib+\frac{4}{3}J^3-
  (\Db H+\D\Hb )\Psi\Psib + \frac{1}{2}\Psi'\Psib\right). \label{ham}
\eea
The flows are defined by
\be \label{eqmot1}
\frac{\partial}{\partial t_k} {\cal A}(Z) = \left\{ {\cal A}(Z) ,
 \int \; dX {\cal H}_k(X) \right\}
\ee
for a given superfield $\cal A$. Since all the Hamiltonians \p{ham}
have vanishing Poisson brackets with $H,\Hb$, we can consistently
put $H=\Hb=0$ in the r.h.s. of \p{eqmot1} {\it after} evaluating the
Poisson brackets. 

The first flow just gives the following equations of motion for 
$J,\Psi,\Psib$
\be\label{triv}
\frac{\partial}{\partial t_1} J =  J' \; , \quad
\frac{\partial}{\partial t_1} \Psi =  \Psi' \; , \quad
\frac{\partial}{\partial t_1} \Psib =  \Psib{}' \; , 
\ee
while the second and third Hamiltonians generate the equations \p{n4kdv2},
\p{n4kdv3}.

It is interesting that the same coset describes another integrable hierarchy
\cite{DGI} which is a extension of $N=2$ KdV
hierarchy with $a=-2$ \cite{ML} and can be defined
by the following Lax operator and Lax equation \cite{DGI}:
\be\label{DGILax}
L= \D\Db\partial +2\D J\Db +\D \Psi \partial^{-1} \Psib \Db \; , \;~~~~~~~~
\frac{\partial}{\partial t_k} L = - 
  \left[ \left( L^{\frac{k}{2}}\right)_{\geq 1} , L \right] \;.
\ee
The first two non-trivial (i.e. the second and third) flow equations of 
motion are
\bea 
\frac{\partial J}{\partial t_2} & = & -\left( \Psib \Psi \right){}' \; ,~~ \;
\frac{\partial \Psi}{\partial t_2}  =  \Psi{}''-
     2\D \left( J \Db\Psi \right) \; ,~~ \;
\frac{\partial \Psib}{\partial t_2}  =  -\Psib{}''-
   2\Db\left( J \D\Psib \right)  \label{DGIeq2}
\eea
and
\bea 
\frac{\partial J}{\partial t_3} & = & J'''+3\left(\DDb J\; J\right)'-
  \frac{3}{2}\left(\DDb J^2\right)'-2\left( J^3\right)'-
   3\left(\Psi\Psib{}'-\Psi'\Psib\right)'+6\left( J\Psi\Psib\right)' \;,\nn
\frac{\partial \Psi}{\partial t_3} & = & 4\Psi{}'''-
  6\D \left( \Db\Psi J'+2\Db\Psi{}' J+J^2\Db\Psi+
  \Psi\Psib\Db\Psi \right)\; ,\nn
\frac{\partial \Psib}{\partial t_3} & = & 4\Psib{}'''+
  6\Db \left( \D\Psib J'+2\D\Psib{}' J-J^2\D\Psib-
  \Psi\Psib\D\Psib \right)\; .\label{DGIeq3}
\eea

Like the previous case the algebra \p{newsca} provides the second
Hamiltonian structure for the hierarchy \p{DGILax} via the coset 
construction. Explicitly, the first three Hamiltonians read 
\bea
H_1 & = & \int dzd\theta d{\bar \theta}  J \; , \quad\quad
H_2  = \int dzd\theta d{\bar \theta} \left(  
  \frac{1}{2}\Psi\Psib \right)\; , \nn
H_3 & = & \int dzd\theta d{\bar \theta} \left(  
    J\DDb J +2\Psi'\Psib+2J\Psi\Psib-\frac{2}{3}J^3-
    4(\Db H+\D\Hb )\Psi\Psib\right), \label{hamm}
\eea
where $\Psi$ and $\Psib$ are subject to the constraints \p{constr} and
the different flow equations are defined by \p{eqmot1}.

Thus both, "fermionic" extensions of $N=2$ KdV hierarchy with $a=4$ 
and $a=-2$ can be treated within the coset approach
as $N=2$ quotients of the $N=2$ nonlinear algebra \p{newsca}.
It is remarkable that although the other details of these two hierarchies are
different, the second hamiltonian structure is the same, as was pointed out 
above.

\setcounter{equation}0
\section{Second Hamiltonian structure}

In the previous section we used the coset formalism applied to the
$N=2$ algebra \p{newsca} to describe the "fermionic" extensions 
of $N=2$ $a=4,-2$ KdV hierarchies. This approach allows for 
a simpler analysis than the Dirac constraint procedure, because the 
constraints $H=\Hb=0$ can be imposed in the equations of motion.
In this Section we will show how to apply the constraints $H=\Hb=0$
to the algebra \p{newsca} and will demonstrate that the Dirac 
brackets for the remaining supercurrents $J,\Psi,\Psib$ close on a
non-local algebra \cite{{DG},{BS}}.

Let us start with the obvious remark that the constraints
\be \label{con1}
H=\Hb=0,
\ee
we want to apply, are incompatible with the Poisson brackets of our
algebra
\be
\left\{ H(1),\Hb (2) \right\} = \frac{1}{2} \D\Db \delta (1,2) \; , \quad
\left\{ \Hb(1), H (2) \right\} =  -\frac{1}{2}\Db\D \delta (1,2) \; , 
  \label{dirac1}
\ee
so we need to apply Dirac's procedure passing to the Dirac brackets
\be \label{db}
\left\{ {\cal A},{\cal B}  \right\}{}^{*} \equiv
\left\{ {\cal A},{\cal B}  \right\}-
\left\{ {\cal A},H  \right\}\left(-\frac{1}{2}\Db\D\right)^{-1}
       \left\{\Hb ,{\cal B}  \right\}-
\left\{ {\cal A},\Hb  \right\}\left(\frac{1}{2}\D\Db\right)^{-1}
      \left\{H,{\cal B}  \right\}
\ee
for any two superfields ${\cal A},{\cal B}$. One can expect some extra
complications in the Dirac procedure because the operators $\D\Db$ and
$\Db\D$ are not invertible on the whole $N=2$ superspace. But 
fortunately this is not so for the case at hand. Indeed, due to the
chiral-anti-chiral structure of $H$ and $\Hb$, the Poisson brackets
$
\left\{ {\cal A},H  \right\}, \left\{\Hb ,{\cal B}  \right\},
\left\{ {\cal A},\Hb  \right\}, \left\{H,{\cal B}  \right\}
$
can be always rewritten in the following form 
\bea
\left\{ {\cal A},H  \right\} & \sim & (\ldots) \D \delta (1,2) \; , \quad
\left\{ {\cal A},\Hb  \right\} \sim (\ldots) \Db \delta (1,2) \; , \nn 
\left\{H ,{\cal B}  \right\} & \sim & \D (\ldots) \delta (1,2) \; , \quad
\left\{\Hb ,{\cal B}  \right\} \sim \Db (\ldots) \delta (1,2) \; , 
\eea
where $(\ldots )$ is used to denote any combinations of operators and/or
superfields.  Thus, one can immediately conclude that we have to know
the operators $(\D\Db)^{-1}$ and $(\Db\D)^{-1}$ not on the whole
$N=2$ superspace but only on its chiral-anti-chiral subspaces, where they
are well defined. Explicitly
\bea
\D \; \left( -\frac{1}{2}\Db\D\right)^{-1} \; \Db & = &  
  \D \; \left( 2\partial^{-1}\right) \; \Db 
  \quad \mbox{because} \quad  
   \D \; \left(-\Db\D\partial^{-1}\right) \; \Db 
                \equiv \D\Db \;, \nn
\Db \;\left(\frac{1}{2}\D\Db\right)^{-1} \; \D & = &  
  \Db \; \left(-2\partial^{-1}\right) \; \D 
  \quad \mbox{because} \quad  
  \Db \; \left(-\D\Db\partial^{-1}\right) \; \D 
                \equiv \Db\D \;.
\eea
Keeping this in the mind all calculations can be done straightforwardly.

All Dirac brackets containing $H,\Hb$ are now equal to zero, so we can
impose the constraints \p{con1} in strong sense.
The rest of the Dirac brackets between the superfields $J,\Psi,\Psib$ form the
following closed non-local algebra
\bea
\left\{ J(1),J(2) \right\}{}^{*}& = & -\left( J\partial +\partial J+
 \D J\Db+\Db J\D+ \frac{1}{2}\partial \DDb \right) \delta (1,2) \; ,\nn
\left\{ J(1),\Psi (2) \right\}{}^{*}& = & \left( 
    -\Db \Psi\D+ \Psi\DDb \right) \delta (1,2) \; ,\nn
\left\{ J(1),\Psib (2) \right\}{}^{*} & = &  \left( 
 -\D \Psib\Db +\Psib\DDb \right) \delta (1,2) \; ,\nn
\left\{ \Psi (1),\Psi (2) \right\}{}^{*}& = & 
    -4\Psi\Db\D\partial^{-1}\Psi \delta (1,2) \; , \;
\left\{ \Psib (1),\Psib (2) \right\}{}^{*} = 
     4\Psib\D\Db\partial^{-1}\Psib \delta (1,2) \; , \nn
\left\{ \Psi (1),\Psib (2) \right\}{}^{*}& = &
 -2\left( \D\Db\partial+2\D J\Db-
    2\Psi\D\Db\partial^{-1}\Psib\right) \delta (1,2)
\label{newscanl}
\end{eqnarray}
with $\Psi,\Psib$ obeying the chirality constraints
\be
\D\Psi=\Db\Psib=0 \;.
\ee
The algebra \p{newscanl} is a particular case of the non-local algebras
proposed in \cite{{DG},{BS}}. Let us note that validity of the
Jacobi identities for the non-local algebra \p{newscanl} is guaranteed
since we constructed it by applying the standard Dirac's procedure applied
to the local algebra \p{newsca}.

Now one can straightforwardly check that the Hamiltonians
\bea
H_1 & = & \int dzd\theta d{\bar \theta}  J \; ,\quad\quad \; 
 H_2  = \int dzd\theta d{\bar \theta} \left( 
 -J^2+\frac{1}{2}\Psib\Psi\right) \;, \nn
H_3 & = & \int dzd\theta d{\bar \theta} \left( 
 J\DDb J +2J\Psi\Psib+\frac{4}{3}J^3+
   \frac{1}{2}\Psi'\Psib\right)     \label{hamd1}
\eea
and
\bea
H_1 & = & \int dzd\theta d{\bar \theta} J \; , \quad\quad\; H_2  = 
 \int dzd\theta d{\bar \theta} \left(  \frac{1}{2}\Psi\Psib\right) \; , \nn
H_3 & = & \int dzd\theta d{\bar \theta} \left(  
 J\DDb J +2\Psi'\Psib+2J\Psi\Psib-\frac{2}{3}J^3 \right)\label{hamd2}
\eea
produce the correct flow equations for the "fermionic" extensions of $N=2$ 
$a=4$ and $a=-2$ KdV hierarchies; they thus correspondingly 
provide the second Hamiltonian structure for both hierarchies.

\setcounter{equation}0
\section{New extensions of N=2 KdV}

In the previous sections we have showed that the second hamiltonian
structure for the "fermionic" extensions of the $N=2$ $a=4$ and $a=-2$
KdV hierarchies in the coset approach
is given by the
very non trivial non-local algebra \p{newscanl}. At the same time
these hierarchies can be obtained via hamiltonian reduction starting from the
local algebra \p{newsca}. One may wonder whether there exist non trivial
integrable hierarchies that include all the superfields $(J,\Psi,\Psib,H,\Hb)$
and possess the algebra \p{newsca} as their second Hamiltonian structure.
This question is rather complicated and will be addressed elsewhere.
Here, instead, we will discuss a simpler problem concerning the existence
of integrable hierarchies with second Hamitonian structure
given by the following {\it linear} subalgebra of \p{newsca}
\bea
\left\{ J(1),J(2) \right\}& = & -\left( J\partial +\partial J+
 \D J\Db+\Db J\D - \frac{1}{2}\partial \DDb \right) \delta (1,2)  ,\nn
\left\{ H(1),\Hb (2) \right\}& = & \D\Db \delta (1,2)  ,
   \label{sca}
\end{eqnarray}
while all the other PB's vanish.  This is the direct product of the $N=2$ SCA 
and the affine ${\hat U}(1)$ algebra. We will show that there exist
at least the first three non-trivial flows, or, in other words
the first four Hamiltonians $(H_1-H_4)$ which have
the proper dimensions and are mutually commuting with respect
to the Poisson brackets \p{sca}. Despite the very trivial Hamiltonian structure
there are three non-equivalent systems for which we succeeded in finding
the first four hamiltonians. The first Hamiltonian is the same for all 
cases $H_1=\int dzd\theta{\bar \theta}(J+H\Hb)$ and corresponds to the 
trivial flow equations, 
like \p{triv}. Due to the very complicated structure of $H_3$ and $H_4$, 
we present here, in addition, only the second hamiltonians for
these systems together with the corresponding second flow equations
(for the explicit form of the third Hamiltonians see Appendix). 
\vspace{0,5cm}\\
\noindent{\bf Extension of $N=2, a= 4$ KdV.}  The first system is given by
the following Hamiltonian density and second flow equations
\be\label{a4}
{\cal H}_2^{(1)}=J^2 +H\Hb{}'+4JH\Hb+\sqrt{2}H\Hb(\Db H + \D\Hb)
\ee
\bea 
\frac{\partial J}{\partial t_2} & = & -\DDb J'-4JJ'+4\Db J\D(H\Hb)+
        4\D J \Db(H\Hb) \nn
 & & +\left( 4JH\Hb-4\Db H\D \Hb+H'\Hb-H\Hb{}'\right)' \; , \nn
\frac{\partial H}{\partial t_2} & = & -H''-
   \D \left( 4\Db (JH)+\sqrt{2}\Db H\Db H+2\sqrt{2}(H\Hb)' \right) \; , \nn
\frac{\partial \Hb}{\partial t_2} & = & \Hb{}''-
   \Db \left( 4\D (J\Hb)+\sqrt{2}\D \Hb\D \Hb-2\sqrt{2}(H\Hb)' \right) 
  \; . \label{a4eq}
\eea
One can immediately see that the system \p{a4eq} admits the reduction 
$H=\Hb=0$. The latter coincides with the $N=2,a=4$ KdV hierarchy.
\vspace{0,5cm}\\
\noindent{\bf Extension of $N=2, a= -2$ KdV.}  The second solution
one can construct is described by the following  flow equations
\bea 
\frac{\partial J}{\partial t_2} & = & \Db J\D(H\Hb)+
        \D J \Db(H\Hb) +
 \left( JH\Hb-\Db H\D \Hb+\frac{1}{2}(H'\Hb-H\Hb{}')\right)' \; , \nn
\frac{\partial H}{\partial t_2} & = & \frac{1}{2}H''-
          \D \left( \Db (JH)+\frac{\sqrt{2}}{2}\Db H\Db H+
          \sqrt{2}(H\Hb)' \right) \; , \nn
\frac{\partial \Hb}{\partial t_2} & = & -\frac{1}{2}\Hb{}''-
   \Db \left( \D (J\Hb)+\frac{\sqrt{2}}{2}\D \Hb\D \Hb-
          \sqrt{2}(H\Hb)' \right) 
  \; , \label{a2eq}
\eea
with 
the Hamiltonian density
\be\label{a2}
{\cal H}_2^{(2)}=
 -\frac{1}{2}H\Hb{}'+JH\Hb+\frac{1}{\sqrt{2}}H\Hb(\Db H + \D\Hb)
\ee
Again the equations \p{a2eq} admit the solution $H=\Hb=0$, which corresponds
to the well--known  $N=2, a=-2$ KdV hierarchy. \vspace{0,5cm}\\

\noindent{\bf New $N=2$ supersymmetric system.}  It comes as a surprise
that there is a third solution, which does not correspond to
any known integrable system. It is described by the following Hamiltonian
density and corresponding second flow equations:
\be\label{new}
{\cal H}_2^{(3)}=
 J^2-3H\Hb{}'+2\sqrt{2}J(\Db H+\D\Hb)+\sqrt{2}H\Hb(\Db H + \D\Hb)
\ee
and
\bea 
\frac{\partial J}{\partial t_2} & = & -\DDb J'+4JJ'-2\sqrt{2}(\Db J H'+
   \D J\Hb{}') \nn
 & & +\sqrt{2}\left( \D\Hb{}'-\Db H'+2J\Db H+2J\D\Hb\right)'\; , \nn
\frac{\partial H}{\partial t_2} & = & 3H''-
          2\sqrt{2}\D \left( J'+\frac{1}{2}\Db H\Db H+
          (H\Hb)' \right) \; , \nn
\frac{\partial \Hb}{\partial t_2} & = & -3 \Hb{}''+
   2\sqrt{2}\Db \left( J'-\frac{1}{2}\D \Hb\D \Hb+
          (H\Hb)' \right) 
  \; . \label{neweq}
\eea
One can see that the system \p{neweq} does not admit the solution $H=\Hb=0$
and so it cannot be reduced, like the previous cases, to a
KdV--type equation on a single superfield $J$.

Of course, the existence of the first four Hamiltonians
for the three systems does not guarantee by itself their
integrability but it is a good reason to take these systems seriously
(i.e. search for Lax operators and Lax equations).

Finally let us stress, that if we take, instead of the Hamiltonian structure
\p{sca}, the direct product of two $N=2$ SCA's (e.g. constructing the second
$N=2$ SCA through the Sugawara procedure from $H,\Hb$) we do not succeed
in finding any system admitting second flows \cite{ZP}.

\setcounter{equation}0
\section{Discussions and outlook}

In this letter we have described the two hierarchies - the
"fermionic" extensions of $N=2$ KdV with $a=4$ \p{ham} and 
$a=-2$ \p{hamm}  within the coset framework. We 
have explicitly showed that both hierarchies can be obtained by quotienting
the $N=2$ superalgebra \p{newsca} by its $N=2$ 
$\widehat {U(1)}$ subalgebra. We also applied Dirac's procedure to
construct the second Hamiltonian structure for these hierarchies.
The resulting $N=2$ superalgebra \p{newscanl} is non-local for these
hierarchies in contrast with the former $N=4$ KdV \cite{DI} and 
"bosonic" extension of the $N=2$ KdV hierarchy \cite{DGI}, which  possess
local linear "small" $N=4$ SCA as second Hamiltonian structure.
We also proposed three additional (presumably) integrable extensions of
the $N=2$ KdV hierarchy which possess at least the four first conserved
currents.

The main advantage of the coset approach is that 
the original $N=2$ superalgebra \p{newsca}
is {\it local}. One may therefore conjecture that also other hierarchies which
can be obtained from the Lax operator \p{laxDGe}
can also be described as a coset of some {\it local} $N=2$ superalgebra.
Moreover, as was shown in \cite{IKT} for the NLS case, these extended 
superalgebras, which localized the second Hamiltonian structure for NLS 
hierarchy \cite{KST}, 
could give rise to new hierarchies with an extended
set of superfields. We defer to a future publication the analysis 
of whether it is possible to construct new integrable hierarchies with 
the superalgebra \p{newsca} as second Hamiltonian structure.

Another important comment concerns the possible existence of other $N=2$ GNLS
hierarchies, besides those constructed in \cite{BKS}. Indeed, while the known
$N=2$ GNLS hierarchy corresponds to the $N=2$ $a=4$ KdV and its
bosonic/fermionic extensions, the existence of the 
"bosonic" and/or "fermionic" extensions of $N=2$ KdV hierarchies \cite{DGI}
with $a=-2$ with the same second Hamiltonian structure seems to imply 
existence of the corresponding "new" GNLS hierarchies. Such "new"
$N=2$ GNLS hierarchies do exist, and will be considered in a
forthcoming paper \cite{BK}. \vspace{0.5cm}

\noindent {\bf Acknowledgments.}
S.K. is grateful to E. Ivanov for many useful discussions.
He also thanks SISSA for the hospitality and financial support extended
to him during the course of this work.

This investigation has been supported in part by the Russian Foundation of
Fundamental Research, grant No. 96-02-17634, RFBR-DFG grant No. 96-02-00180,
INTAS grant 94-2317, and by a grant from the Dutch NWO organization.

\setcounter{equation}0\section*{Appendix}
\def\theequation{A.\arabic{equation}}

Here we write down the expressions for the third Hamiltonians for the
systems \p{a4eq}, \p{a2eq}, 
\p{neweq}.
The general expression for the third Hamiltonian density have the 
following form

\begin{eqnarray}
{\cal H}_3 &= &
  a_{1 }(\DDb J)J +a_{2 } J\Db H'+a_{3 }J\D\Hb{}' + 
  a_{4 }H\Hb{}''+a_{5 }(\Db H') H\Hb  + 
  a_{6 }(\DDb J)H\Hb \nn
 & & +a_{7 }H(\D\Hb{}')\Hb  + a_{8 }(\Db J)(\Db H)H + a_{9 }(\Db J)JH  + 
  a_{10} (\D J)(\D\Hb)\Hb + a_{11} (\D J)J\Hb  \nn
 & &  +   a_{12} J(\Db H)(\D\Hb) 
  + a_{13} J^3 + a_{14} (\Db H)^2 H\Hb  + a_{15} (\Db H)H(\D \Hb)\Hb  \nn
 & & +  a_{16} (\D \Hb)^2 H\Hb   +   a_{17} J(\Db H) H\Hb  + 
  a_{18} J(\D \Hb)H \Hb  + a_{19} J^2 H\Hb . \label{A1}
\end{eqnarray}
The solutions we obtained are
\vspace{0.5cm} \\
\begin{tabular}{|r|r|r|r|r|r|r|r|r|r|} \hline \hline
    Case & $a_1$ & $a_2$ & $a_3$ & $a_4$ & $a_5$       & $a_6$ &   $a_7$   & 
          $a_8$ & $a_9$  \\ \hline
 a=4 KdV &   1   &  0    &   0   &  -1   & $3\sqrt{2}$ &0& $-3\sqrt{2}$&  
          0    &  0    \\
a=-2 KdV & -1/6  &  0    &   0   &  2/3  & $2\sqrt{2}$ &  -1   & 
         $-2\sqrt{2}$&   0    &  0     \\
new  KdV &  1/3  & $1/\sqrt{2}$& $-1/\sqrt{2}$ & 2/3 & $3/(8\sqrt{2})$ & 
         1 & $-3/\sqrt{2}$ &-1& $-\sqrt{2}$ \\ \hline \hline
$a_{10}$& $a_{11}$ & $a_{12}$ & $a_{13} $&$a_{14}$ & $a_{15}$ & $a_{16}$ &
         $a_{17}$ & $a_{18}$ & $a_{19}$ \\ \hline
  0  & 0 & 6 & -4/3 &  -
2 & 0  & -2 & $-6\sqrt{2}$   & $-6\sqrt{2}$  & -12 \\
     0  & 0 & 2 & -1/9 & 4/3 & 3 & 4/3 & $2\sqrt{2}$ &$2\sqrt{2}$ & 1 \\
    -1  & $-\sqrt{2}$  & 1 & 2/9 & 1/3 & 1 & 1/3 & 0 & 0 &  0 \\
\hline \hline
\end{tabular}
\vspace{0.5cm}\\

\end{document}